\documentclass[hidelinks]{article}

\usepackage{arxiv}

\usepackage[utf8]{inputenc} 
\usepackage[T1]{fontenc}    
\usepackage{hyperref}       
\usepackage{url}            
\usepackage{booktabs}       
\usepackage{amsfonts}       
\usepackage{nicefrac}       
\usepackage{microtype}      
\usepackage{lipsum}		
\usepackage{graphicx}
\usepackage{natbib}
\usepackage{doi}

\usepackage[detect-all,range-units=single]{siunitx}
\usepackage{siunitx}
\usepackage[super]{nth}
\usepackage{textcomp}
\usepackage{booktabs}
\usepackage[table]{xcolor}

\title{Experimental Study of the Phase Noise in K-band ARoF systems for Low Complexity 5G receivers}

\author{ \href{https://orcid.org/0000-0003-4444-7157}{\includegraphics[scale=0.06]{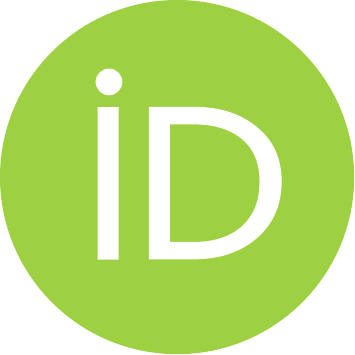}\hspace{1mm}Javier~Pérez~Santacruz}\\
	Institute for Photonic Integration\\
	Eindhoven University of Technology\\
	5600~MB Eindhoven, the Netherlands \\
	\texttt{j.perez.santacruz@tue.nl} \\
	\And
	\href{https://orcid.org/0000-0001-8279-8180}{\includegraphics[scale=0.06]{figures/orcid.pdf}\hspace{1mm}Simon~Rommel} \\
	Institute for Photonic Integration\\
	Eindhoven University of Technology\\
	5600~MB Eindhoven, the Netherlands \\
	\texttt{s.rommel@tue.nl} \\
	
	\And
	\href{https://orcid.org/0000-0002-6498-0699}{\includegraphics[scale=0.06]{figures/orcid.pdf}\hspace{1mm}Ulf~Johannsen} \\
	Centre for Wireless Technology\\
	Eindhoven University of Technology\\
	5600~MB Eindhoven, the Netherlands \\
	\texttt{u.johannsen@tue.nl} \\
	
    \And
	\href{https://orcid.org/0000-0002-4381-5109}{\includegraphics[scale=0.06]{figures/orcid.pdf}\hspace{1mm}Antonio~Jurado-Navas} \\
    Communication Engineering Department\\
	University of Málaga\\
	29071~Málaga, Spain \\
	\texttt{u.johannsen@tue.nl} \\
	
	\And
	\href{https://orcid.org/0000-0002-2935-7682}{\includegraphics[scale=0.06]{figures/orcid.pdf}\hspace{1mm}Idelfonso~Tafur~Monroy} \\
	Institute for Photonic Integration\\
	Eindhoven University of Technology\\
	5600~MB Eindhoven, the Netherlands \\
	\texttt{i.tafur.monroy@tue.nl} \\
    
}

\renewcommand{\shorttitle}{\textit{arXiv} Template}

\hypersetup{
pdftitle={A template for the arxiv style},
pdfsubject={q-bio.NC, q-bio.QM},
pdfauthor={David S.~Hippocampus, Elias D.~Striatum},
pdfkeywords={First keyword, Second keyword, More},
}

\begin{document}
\maketitle

\begin{abstract}
	In this paper, an experimental analysis of the phase noise in an OFDM ARoF setup at \SI{25}{GHz} for beyond 5G is presented. The configuration of the setup allows to gradually scale the final phase noise level of the system. Moreover, an OFDM phase noise mitigation method with low complexity and delay is proposed and explained. The proposed method is an advanced version of the LI-CPE algorithm. The advanced LI-CPE version avoids the one OFDM symbol delay of its antecedent. In addition, the yields of using both methods are shown under different phase noise levels and with different subcarrier spacings. Finally, it is experimentally proven that the proposed method performs better than its previous version.
\end{abstract}

\keywords{ARoF \and OFDM \and phase noise \and 5G}

\section{Introduction}
The exponential growth of the mobile network data has motivated to research about new technologies that can support future traffic requests. The fifth-generation (5G) of wireless networks is the established solution to these demanding requirements. Moreover, new types of services with multiple requirements are emerging, and the mobile network has to adapt to them. Therefore, 5G proposes three types of scenarios to fulfill such services\cite{3GPP_release15}: enhanced mobile broadband (eMBB) to high bit rate services; massive machine-type communications (mMTC) to support a huge quantity of low power connected devices; and ultra-reliable and low latency communications (URLLC) where latency and reliability are the priorities.

One of the most prominent ways to increase the bit rate is moving from the current saturated band to higher frequencies, in the millimeter-wave (mm-wave) domain. Furthermore, analog radio-over-fiber (ARoF) is a suited technology due to its long reach distances, low cost, wide bandwidth, high spectral efficiency, and low power consumption\cite{mmWave_5G}. In addition, the centralized-radio access network (C-RAN) is a trending architecture because it offers attractive benefits such as flexibility, low latency, and reduced energy consumption. Hence, mm-wave ARoF over a C-RAN scheme is a strong candidate to be part of the future 5G structure. However, mm-wave ARoF presents several drawbacks such as phase noise, chromatic dispersion, nonlinearities, or high free-space path loss (FSPL).

Orthogonal frequency division multiplexing (OFDM) is the modulation format chosen by the 3GPP 5G standard~\cite{3GPP_release15}. OFDM communications bring advantages like robustness to frequency selective channels, high spectral efficiency, or efficient multiple-input multiple-output (MIMO) integration. Nonetheless, one of the major OFDM degrading impairments is the phase noise. Thence, a phase noise compensation algorithm has to be performed in an mm-wave OFDM ARoF system. There are three types of OFDM phase noise mitigation techniques~\cite{phase_noise_types}: decision-feedback-based schemes, blind estimation schemes, and pilot-based schemes. Decision-feedback and blind estimation schemes are complex iterative methods. Moreover, these iterative algorithms show difficulties to converge in some cases. Then, these types of techniques are not suitable for mMTC and URLLC scenarios. Pilot-based schemes are appropriate phase noise compensation methods for these 5G scenarios due to its simplicity and low latency process.

In this work, the phase noise degradations are studied and analyzed in an experimental OFDM ARoF setup at \SI{25}{GHz} (K-band). The setup scheme can increment gradually the final phase noise of the received signal in the system. Furthermore, an OFDM phase noise mitigation algorithm with low complexity and latency is proposed. This algorithm is compared with another method and probed experimentally. The proposed method is analyzed under different OFDM configurations and with different phase noise levels.

\section{Phase noise effect on OFDM systems and proposed algorithm to compensate it}
\label{sec: OFDM_algo}

Phase noise causes two degradations in OFDM signals. The first one is a common phase error (CPE) that affects all the subcarriers in the same way and can be compensated by performing the zero-forcing (ZF) equalizer. The second degradation consists of inter-carrier interference (ICI). ICI can not be mitigated by a normal equalizer. Hence, an additional method has to be included to compensate for the ICI caused by the phase noise. The OFDM symbol duration plays an important role in a phase noise channel. This fact is because the ICI originated by the phase noise is proportional to the OFDM symbol period. The subcarrier spacing is inversely proportional to the OFDM symbol duration. Then, the phase noise degradation decreases with the subcarrier spacing in OFDM systems.

Furthermore, most of the phase noise comes from the oscillators in a communication system and follows a Wiener process~\cite{wiener_phase}. A very simple method to estimate the phase noise in this type of scenario consists of the interpolation of the CPEs that belong to consecutive OFDM symbols~\cite{interp_method}. This method is based on the CPE represents the middle phase noise point of the OFDM symbol in the time domain. A way to estimate the CPE by using pilot subcarriers is~\cite{interp_method}:

\begin{equation}
\label{eq: CPE}
{\bf CPE}^{m} = \text{angle} \left ({\bf Y}^{m}_{k} \times \text{conj}({\bf H}^{m}_{k} \cdot  {\bf X}^{m}_{k})\right ), \;\;  k \; \epsilon \; {\bf PSC}  
\end{equation}

\noindent where ${\bf Y}$ is the received symbols after the fast Fourier transform (FFT) process; ${\bf H}$ represents the estimated channel in the frequency domain; ${\bf X}$ is the vector of the transmitted subcarriers; $k$ is the subindex of the pilot subcarriers (${\bf PSC}$) in the vectors ${\bf Y}$, ${\bf H}$, and ${\bf X}$; the operators $\times$ and $\cdot$ determine a vector multiplication and a multiplication element by element, respectively; the hyperindex $m$ concerns with the OFDM symbol index. After this CPE estimation, the phase noise can be linear interpolated and compensated. This method is called linear interpolation based ICI estimation technique (LI-CPE)~\cite{interp_method}. The main advantages of LI-CPE are its low complexity and its robustness because a noniterative process is implemented. However, applying LI-CPE supposes a delay of one OFDM symbol since it is necessary to calculate the CPE of the next OFDM symbol for the phase noise estimation of the current OFDM symbol. 

The proposed OFDM mitigation algorithm is an advanced technique of the LI-CPE method. This advance is based on using the redundancy of the OFDM structure where the prefix cyclic (CP) is a copy of the last part of the OFDM symbol. It is possible to estimate the phase noise slope in each OFDM symbol by processing the CP and the last part of the symbol. This slope can be roughly calculated by employing the next formula: 

\begin{equation}
\label{eq: slope}
  \Delta {\bf\phi}^{m} = \left (  \sum_{i=1}^{N_{cp}} \text{angle} \left ( \frac{{\bf r}_{i+N}^{m}}{{\bf r}_{i}^{m}}  \right ) \right ) \Biggm/ \left ( N_{cp}\cdot N\right )
\end{equation}

\noindent where $N$ is the total number of subcarriers; $N_{cp}$ is the number of CP samples; ${\bf r}$ is the received signal in the time domain (including the CP). Therefore, a linear interpolation of the phase noise can be performed by using the CPE and this slope. So, the phase noise estimation is reached although the next linear equation:

\begin{equation}
    {\bf\widehat{\rho }}_{\,k}^{\,m} = \Delta {\bf\phi}^{m} \cdot \left ( k - \frac{N_{T}}{2} \right ) + {\bf CPE}^{m}, \;\;  k \; \epsilon \; [0, N_{T}-1]
\end{equation}

 \noindent where $N_{T}$ is the sum of $N$ and $N_{cp}$. In this way, the delay of one OFDM symbol is avoided because the phase noise can be independently estimated in each symbol. Moreover, the phase noise estimation is more accurate than in the normal LI-CPE because the interpolation process has more reference points. This idea to estimate the phase noise was initially used in~\cite{iterp_method_slope}.

\begin{figure*}[t]
\centering
\includegraphics[width=1\linewidth]{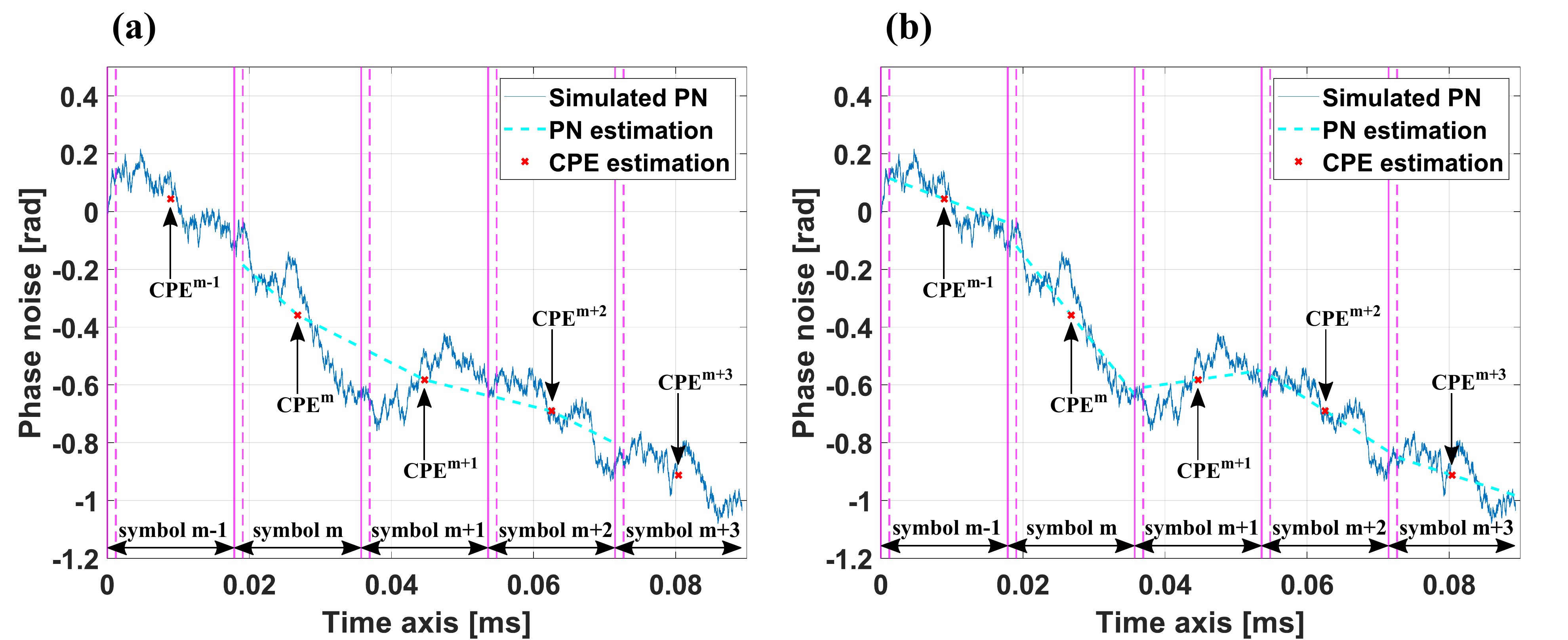}
\caption{Phase noise interpolation based on CPE: LI-CPE in~\cite{interp_method} (a) and proposed LI-CPE (b).}
\label{fig: PN_rep}
\end{figure*}

Fig.~\ref{fig:  PN_rep} shows two graphs to visually see and compare the LI-CPE method proposed in~\cite{interp_method} (Fig.~\ref{fig:  PN_rep}~(a)) and the advanced version of LI-CPE explained in this section (Fig.~\ref{fig:  PN_rep}~(b)). These graphs have been obtained through simulations. The simulated phase noise is created according to a Wiener process~\cite{wiener_phase} with a two-sided 3-dB bandwidth ($\beta$) of \SI{150}{Hz}.  Additive white Gaussian noise (AWGN) and fading channel are not included to see more clearly the phase noise estimation performance of each algorithm. The relevant OFDM parameters used in these simulations are the following: \SI{60}{kHz} of subcarrier spacing, \SI{1.2}{\micro s} of CP period, the total duration of an OFDM symbol is \SI{17.87}{\micro s}, 4096 active subcarriers, and one pilot tone inserted on every \nth{12} active subcarrier.

The ICI level introduced by the phase noise is directly related to the ratio between the OFDM symbol duration and the two-sided 3-dB bandwidth of the phase noise. Moreover, it should be noted that the CPE estimation improves with the number of pilots (see equation~\ref{eq: CPE}). In addition, the phase noise affects more to the lower frequency subcarriers due to its low-pass spectrum nature.

In both graphs, the blue line represents the simulated phase noise to estimate; the dotted cyan plot corresponds to the estimated phase noise; the red crosses are the estimated CPEs; and the vertical solid purple lines show the borders between each OFDM symbol. The period between the vertical purple solid and dotted lines represents the CP duration. The LI-CPE method lacks the first and last phase noise estimation because it is necessary to know the previous and next CPE values to determine the phase noise in each OFDM symbol. Observing the two graphs in Fig.~\ref{fig:  PN_rep},  it can be seen that the phase noise interpolation of the proposed LI-CPE method is better suited. For this example, the mean square error (MSE) values of the phase noise estimation are 0.0106 and 0.0037 for the normal LI-CPE and the advanced LI-CPE, respectively. 

\section{Experimental setup}
\begin{figure*}[t]
\centering
\includegraphics[width=1.0\linewidth]{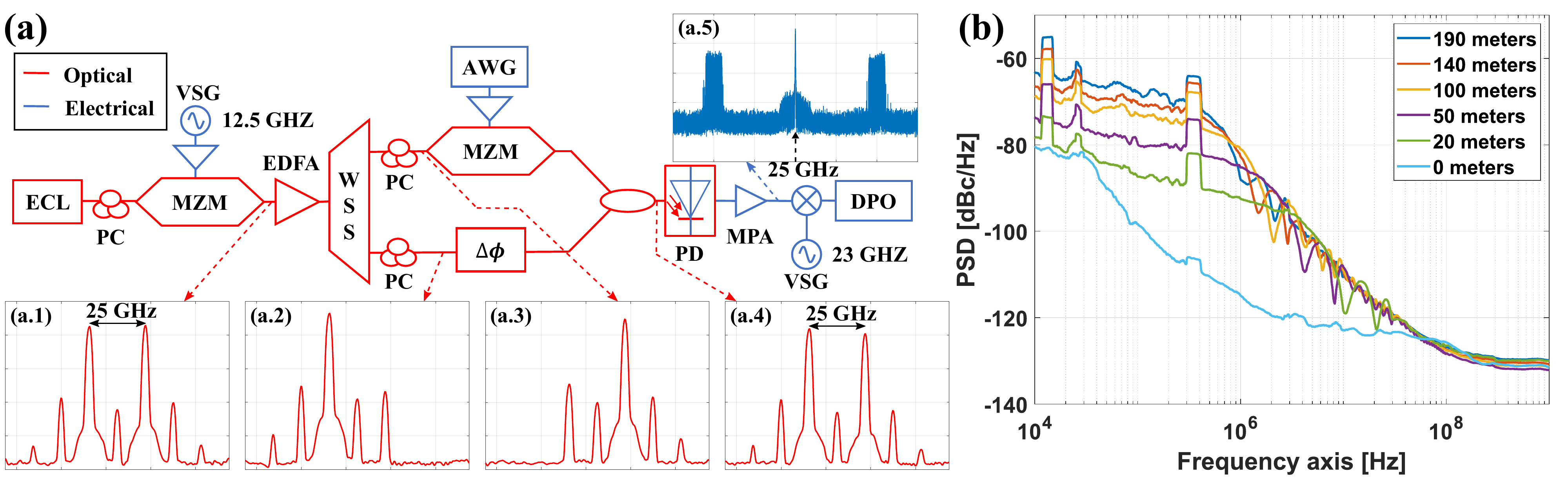}
\caption{Experimental mm-wave ARoF setup (a) and PSD of the measured phase noise with different path lengths.}
\label{fig: setup}
\end{figure*}

The experimental testbed scheme is shown in~Fig.~\ref{fig: setup}~(a). The main purpose of this setup is to adjust the phase noise level in an mm-wave ARoF system. Then, an analysis of the phase noise and its degradations in OFDM signals can be realized. First, an electrical cavity laser (ECL) generates an optical carrier at \SI{1550}{nm} of wavelength. Next, a Mach-Zehnder modulator (MZM), biased in the null point, modulates the optical carrier with an RF carrier of \SI{25}{GHz}. This RF carrier is produced by a vector signal generator (VSG). Therefore, two optical tones, corresponding to the second harmonics of the MZM, are generated with a frequency separation of \SI{25}{GHz}. The spectrum of these two tones are exhibited in~Fig.~\ref{fig: setup}~(a.1).  After the laser, a polarization controller (PC) is utilized because MZMs are sensitive to optical polarization. Then, the optical signal is boosted by an erbium-doped fiber amplifier (EDFA). 

The next step consists of splitting the two tones through a wavelength switch selector (WSS). The graphs~(a.2) and~(a.3) of~Fig.~\ref{fig: setup} show the optical spectrum of the split tones after the filter process. The tone from the upper branch is modulated with the OFDM signal through a second MZM. The OFDM signal is produced by an arbitrary waveform generator (AWG) of \SI{12}{GSa/s}. The other tone is delayed regarding the modulated tone by a patch core in the lower branch. This delay is the key to the scheme to increment gradually the phase noise. The final phase noise level of the system is proportional to the delay between the two branches~\cite{phase_equation}. The modulated and delayed tones are recombined in a coupler. The optical spectrum of this combination is shown in~Fig.~\ref{fig: setup} (a.4). It is relevant to mention that, to get the maximum power in the coupler, the polarization of both tones has to be the same. For this reason, one polarization controller is employed in each branch.

In the photodiode (PD), the recombined tones are beaten and converted to an RF signal of \SI{25}{GHz}. Next, the electrical signal is boosted by a \SI{30}{dB} medium power amplifier (MPA). Fig.~\ref{fig: setup}~(a.5) exhibits the electrical spectrum at the MPA output. The OFDM bands and the \SI{25}{GHz} carrier can be observed in this spectrum. Then, the electrical signal is mixed with a \SI{23}{GHz} sinusoid produced by a second VSG. Consequently, the signal is downconverted to an intermediate frequency (IF) at \SI{2}{GHz}. Finally, the IF signal is sampled by a digital phosphor oscilloscope (DPO) of \SI{12.5}{GSa/s}.

The power spectral density (PSD) of the phase noise is measured before the DPO for several delay values: \SIlist{0;96;240;480;672;912}{ns}. These measures are shown in~Fig.~\ref{fig: setup}~(b). The \SI{0}{meters} case corresponds to the lowest phase noise level. As the path length increases, the PSD phase noise also increments. For path lengths higher than \SI{20}{meters}, the phase noise depicts a fading pattern in the PSD. This fact is because the two branches of the scheme start to perform as an interferometer from \SI{20}{meters} of delay. The PSD phase noise can be calculated through the equations presented in~\cite{phase_equation}. 


\definecolor{LightGray}{gray}{0.95}
\newcolumntype{g}{>{\columncolor{LightGray}}c}

\begin{table}[t]
\centering
\caption{OFDM configuration parameters.}
\label{tab: paprameter_table}
\begin{tabular}{l  ccccc}
\toprule
\textbf{Config.} 
& \textbf{1} & \textbf{2} & \textbf{3}   & \textbf{4} & \textbf{5}  \\ 
\midrule

\rowcolor{LightGray}  $\Delta f$ [KHz] 
& 30 & 60  & 120 & 240 & 480   \\ 

 $N$
 & $ \text{2}^{13}$    & $ \text{2}^{12}$  & $ \text{2}^{11}$    & $ \text{2}^{10}$ &  $\text{2}^{9}$      \\ 

\rowcolor{LightGray}  $T_{cp}$ [\SI{}{\micro s}]
  & 2.4  & 1.2  & 0.6  & 0.3 & 0.15     \\ 

\bottomrule
\end{tabular}
\end{table}

Different OFDM configurations are transmitted in the experimental setup. The common parameters of all the configurations used are: quadrature phase-shift keying (QPSK) modulation, one pilot tone inserted on every \nth{12} active subcarrier, and 80\% of all subcarriers are active. The rest of the parameters are shown in~Table~\ref{tab: paprameter_table} following this order: subcarrier spacing ( $\Delta f$), total number of carriers ($N$), and CP period ( $T_{cp}$). The configuration two is according to the 3GPP 5G standard~\cite{3GPP_release15}. The parameters of the remaining configurations are proportional to this configuration for a fair comparison (same bandwidth and bit rate).

\section{Results}

\begin{figure*}[t]
\centering
\includegraphics[width=1\linewidth]{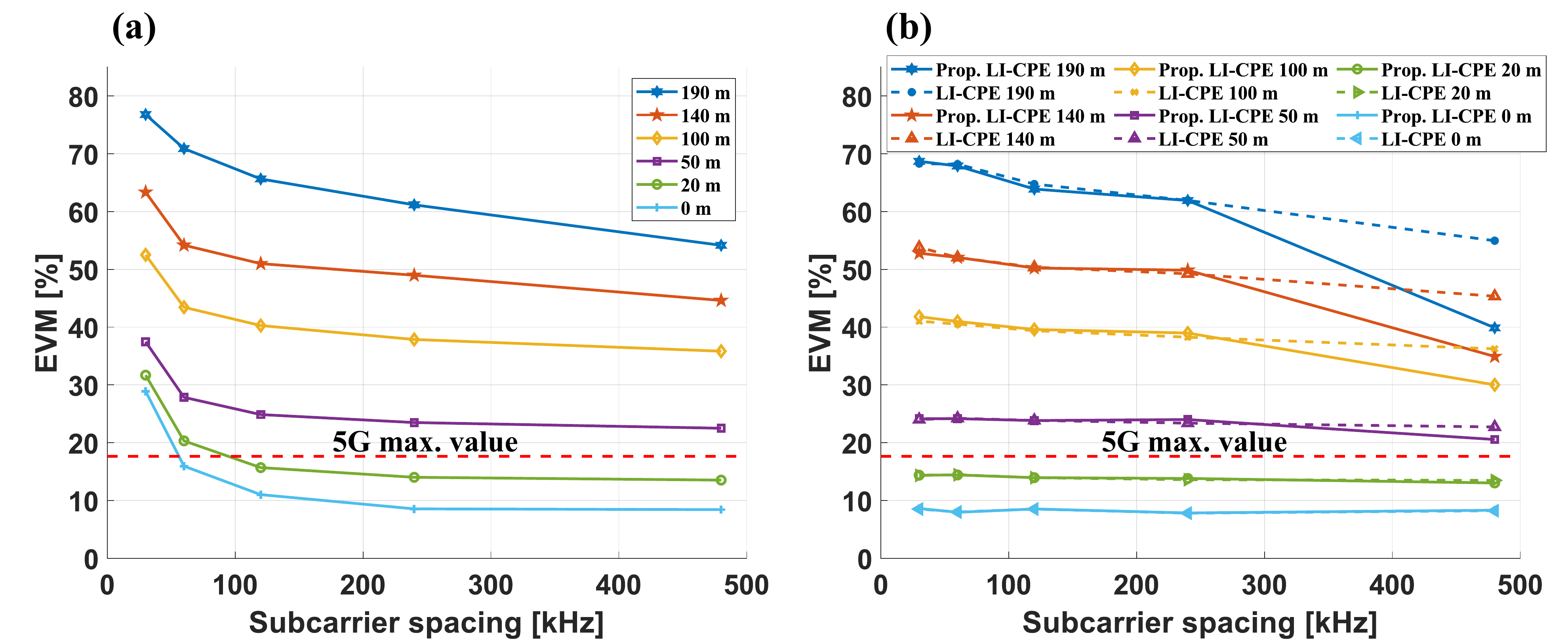}
\caption{EVM as a function of the subcarrier spacing for several path length values: using the standard OFDM receiver (a); adding the LI-CPE method of~\cite{interp_method} and the proposed LI-CPE to the standard OFDM receiver (b). }
\label{fig: evm_after}
\end{figure*}

Fig.~\ref{fig: evm_after} exhibits the experimental results of the previous section. These results are represented in terms of error vector magnitude (EVM) in percent.  The horizontal axis of both graphs depicts the different OFDM configurations in~ Table~\ref{tab: paprameter_table}. Fig.~\ref{fig: evm_after}~(a) corresponds to the EVM by employing the standard OFDM receiver. The standard OFDM receiver consists of the following blocks in this order: remove the CP, FFT, ZF equalizer, and QPSK demodulator. The graph of~Fig.~\ref{fig: evm_after}~(b) adds the LI-CPE methods  explained in~section~\ref{sec: OFDM_algo} before the ZF equalizer : proposed LI-CPE technique (solid line) and LI-CPE algorithm of~\cite{interp_method} (dotted line). The maximum 5G EVM value of 17.5 \% for QPSK modulations is also represented as a dotted red line in both graphs~\cite{3GPP_release15}. Observing Fig.~\ref{fig: evm_after}~(a), the EVM is higher for longer path delays (see~Fig.~\ref{fig: setup}~(b)). In addition, for shorter subcarrier spacing, the EVM is also higher because the OFDM symbol period is larger and the phase noise degrades more the signal. 

On the other hand, the EVM also increments with the path length in Fig.~\ref{fig: evm_after} (b). Nonetheless, the EVM curves for the different delays are flatter than in Fig.~\ref{fig: evm_after}~(a). These flat curves mean that the OFDM symbol period is less related to the ICI introduced by the phase noise if one of the LI-CPE methods is performed. Therefore, the EVM improvement of applying these methods is higher for shorter subcarrier spacing values. Moreover, the EVM values of the proposed LI-CPE are lower for higher subcarrier spacing configurations because the phase noise is better estimated as it was mentioned in section~\ref{sec: OFDM_algo}. Thence, these results prove that the proposed LI-CPE has two big advantages regarding the LI-CPE method presented in~\cite{interp_method}: one OFDM symbol delay less and better EVM values. The only minor disadvantage is the added process to calculate the phase noise slope through equation~\ref{eq: slope}. Nevertheless, this calculation requires a few extra operations. Furthermore, for delay lengths above \SI{20}{meters}, the EVM values exceed the 17.5 \% requirement limit of 5G standard. Hence, a more complex phase noise mitigation algorithm has to be used in these conditions. 

\section{Conclusions}
An analysis of the phase noise has been shown in an experimental ARoF setup for K-band (\SI{25}{GHz}). The importance of employing low complex and latency phase noise mitigation techniques in this type of system for mMTC and URLLC scenarios has been explained too. Moreover, an OFDM phase noise mitigation method with low complexity and latency has been proposed. This method is an advanced version of the LI-CPE technique. Both methods have been studied and compared for different phase noise conditions and subcarrier spacing in the experimental ARoF setup. The results prove that the proposed method outperforms LI-CPE in terms of EVM and, in addition, supposes one OFDM symbol delay less. 

\section*{Acknowledgements}

This work has been partially supported by 5G STEP FWD (GA no.~722429) and blueSPACE (GA no.~762055) projects which have received funding from the European Union\textquotesingle s Horizon 2020 research and innovation programme.

\bibliographystyle{unsrtnat}


\begin{thebibliography}{2}

\bibitem[\protect\citeauthoryear{}{2019}]{3GPP_release15}{3rd Generation Partnership Project; Technical Specification Group Services and System Aspects; Rel. 15 Description; Summary of Rel-15 Work Items (Rel. 15), Feb. 2019.}

\bibitem[\protect\citeauthoryear{}{2019}]{mmWave_5G}{S. Rommel \textit{et al}., "High-Capacity 5G Fronthaul Networks Based on Optical Space Division Multiplexing," \textit{IEEE Trans. Broadcast.}, vol. 65, no. 2, pp. 434-443, Jun. 2019.}

\bibitem[\protect\citeauthoryear{}{2017}]{phase_noise_types}{P. Mathecken \textit{et al}., "Constrained Phase Noise Estimation in OFDM Using Scattered Pilots Without Decision Feedback," in IEEE Trans. Signal Process, vol. 65, no. 9, pp. 2348-2362, May. 2017.}

\bibitem[\protect\citeauthoryear{}{2000}]{wiener_phase}{A. Demir \textit{et al}., "Phase noise in oscillators: a unifying theory and numerical methods for characterization," in IEEE Trans. Circuits Syst. I, Fundam. Theory Appl., vol. 47, no. 5, pp. 655-674, May. 2000.}

\bibitem[\protect\citeauthoryear{}{2011}]{interp_method}{M. E. Mousa-Pasandi and D. V. Plant, "Noniterative Interpolation-Based Partial Phase Noise ICI Mitigation for CO-OFDM Transport Systems," in IEEE Photon. Technol. Lett., vol. 23, no. 21, pp. 1594-1596, Nov. 2011.}

\bibitem[\protect\citeauthoryear{}{2013}]{iterp_method_slope}{Y. Ha and W. Chung, "A Feedforward Partial Phase Noise Mitigation in the Time-Domain using Cyclic Prefix for CO-OFDM Systems," in J. Opt. Soc. Korea  17, pp. 467-470, 2013.}

\bibitem[\protect\citeauthoryear{}{2012}]{phase_equation}{T. Shao \textit{et al}., "Investigation on the Phase Noise and EVM of Digitally Modulated Millimeter Wave Signal in WDM Optical Heterodyning System," in J. Light. Technol., vol. 30, no. 6, pp. 876-885, Mar. 2012.}

\bibitem[\protect\citeauthoryear{}{2008}]{mmWave_tech}{N. Mohamed \textit{et al}., "Review on system architectures for the millimeter-wave generation techniques for RoF communication link,"in Proc. IEEE Int. RF Microw. Conf. (RFM), pp. 326-330, Dec. 2008.}


\end{thebibliography}

\end{document}